\documentclass[pra,12pt,tightenlines]{revtex4}
\usepackage{latexsym}
\usepackage{graphicx}
\usepackage{amsmath}
\usepackage{amsfonts}
\usepackage{amsbsy}
\usepackage{amsthm}
\usepackage{bbm}
\usepackage{bm}
\usepackage{dsfont} 

\newcommand{\A}{\mathcal{A}}

\newcommand{\Ha}{\mathcal{H}}

\begin{document}

\title{Time-ordering Dependence of Measurements in Teleportation}

\author{Reinhold A. Bertlmann} \email{reinhold.bertlmann@univie.ac.at}
\affiliation{University of Vienna, Faculty of Physics, Boltzmanngasse 5, A-1090 Vienna,
Austria}
\author{Heide Narnhofer} \email{heide.narnhofer@univie.ac.at}
\affiliation{University of Vienna, Faculty of Physics, Boltzmanngasse 5, A-1090 Vienna,
Austria}
\author{Walter Thirring} \email{walter.thirring@univie.ac.at}
\affiliation{University of Vienna, Faculty of Physics, Boltzmanngasse 5, A-1090 Vienna,
Austria}


\begin{abstract}

We trace back the phenomenon of ``delayed-choice entanglement swapping'' as it was realized in a recent experiment to the commutativity of the projection operators that are involved in the corresponding measurement process. We also propose an experimental set-up which depends on the order of successive measurements corresponding to noncommutative projection operators. In this case entanglement swapping is used to teleport a quantum state from Alice to Bob, where Bob has now the possibility to examine the noncommutativity within the quantum history.

\end{abstract}

\maketitle


\section{Introduction}\label{sec:introduction}

Recently the Zeilinger group \cite{Ma-Zeilinger} has performed an experiment that has been stimulated by Asher Peres \cite{Peres00}. It was named ``delayed-choice entanglement swapping'' and was perceived as a quantum mechanical paradox if -- as Peres pointed out -- a simple quantum rule is forgotten: ``It is meaningless to assert that two particles are entangled without specifying in which state they are entangled ... or if we attempt to attribute an objective meaning to the quantum state of a single system''. Then quantum effects may mimic an influence of future actions on past events.

However, it always depends on the point of view what will be considered as a paradox. In this article we want to interpret, on one hand, the experimental set-up such that the result appears quite natural even to a classically educated physicist. The paradox arises when the point of view is changed within an interpretation, in particular, when classical probabilities are interpreted as quantum correlations that arise due to a coherent superposition of quantum states. On the other hand, it is essential in the interpretation of the experiment that a measurement in quantum theory always causes a collapse of the quantum state (or wave function), see von Neumann \cite{vonNeumann}. In contrast, a measurement in classical physics we have to interpret as a kind of filtering. By filtering we sort out those states from the set of states that do not possess our desired properties (or boundary conditions). Of course, we also can consider a measurement in quantum theory in such a way. The difference, however, is that in classical physics due to the commutativity of multiplication the order of successive measurements does not matter at all. Whereas in quantum physics (formulated by \emph{quantum histories} \cite{Gell-Mann_Hartle, Griffiths}) we can only be sure that the results do not depend on the order of the measurements if the corresponding projection operators commute. But precisely such (commuting) projection operators enter in the experiment of the Zeilinger group \cite{Ma-Zeilinger}.

Entanglement swapping offers the possibility to an external observer, called Victor, who has access to a Hilbert space which is tensorized with an other Hilbert space, to change a quantum state in the other Hilbert space by performing a measurement in his space. The quantum state, which previously appears separable for Alice and Bob, but not pure, is changed into a new state that is now pure and entangled for Alice and Bob. This new (entangled) state is generated by a collapse of the quantum state. That is only possible for states where Victor is entangled with the total system of Alice and Bob.

However, if this entanglement with Victor gets destroyed by a measurement of Alice and/or Bob -- as it is the case in the experiment of Ref.~\cite{Ma-Zeilinger} -- then also Victor, by performing a measurement, has no chance to deliver an entangled state to Alice and Bob. The correlations that remain can be explained by a mixed state with only classical correlations.

We are interested in entanglement swapping since it gives Alice and Bob the chance for quantum teleportation. This possibility is automatically supplied when Victor performs entanglement swapping. In this case a kind of delayed-choice entanglement swapping is indeed possible as we will explain in this article. Alice may perform first her required measurements and only afterwards Victor establishes the connection between Alice and  Bob. However, the measurements Alice performs and compares with those of Bob are in this delayed-choice experiment not so much related to entanglement swapping but rather to double teleportation. In any case, there is a sequence of measurements where the order of successive measurements does not matter.

On the contrary, there is the possibility that Alice and Victor perform certain measurements that imply, dependent on the measurement results, that Bob receives some well-defined quantum state. Which one, however, depends on the order of successive measurements that now correspond to noncommutative projection operators. By inspection of Bob's quantum state the experimenter has therefore the possibility -- at least in principle -- to examine in a precise way (that means we do not average over the probabilities of the outcomes of the measurements) this noncommutativity within the quantum history.

To demonstrate this noncommutative feature of measurements within the quantum history, we use the mathematical formalism of isometries (introduced already in Refs.~\cite{Uhlmann01, TBKN}) as mappings from one factor to the other in the tensor product of Hilbert spaces. It turns out that this general formalism, being valid in any dimensions, is quite powerful and convenient to handle.

\section{Physical Settings and Formalism}\label{sec:physical settings formalism}

The physical settings we consider are that of quantum teleportation and entanglement swapping. Both are related to a kind of \emph{quantum transport} between subsystems. We consider a tensor product of two Hilbert spaces $\Ha_1 \otimes \Ha_2$ of equal dimensions. Then a projection in $\Ha_1$ reduces a given pure state in $\Ha_1 \otimes \Ha_2$
\begin{equation}\label{state in H1xH2}
|\,\psi\,\rangle_{12} \;=\;\sum\limits_{i=1}^d \,a_i\,|\,\varphi_{i} \,\rangle_1 \otimes |\,\kappa_{i} \,\rangle_2
\end{equation}
to a pure separable state
\begin{equation}\label{projection in H1}
\big(\left|\,\chi\,\right\rangle\left\langle\,\chi\,\right|\big)_1 \otimes \mathds{1}_{2} \, |\,\psi\,\rangle_{12} \;=\; |\,\chi\,\rangle_1 \otimes \sum\limits_{i=1}^d \,a_i\,\left\langle \,\chi\,|\,\varphi_{i} \,\right\rangle_1 |\,\kappa_{i} \,\rangle_2 \,,
\end{equation}
which by appropriate choice of $|\,\chi\,\rangle_1$ and $|\,\psi\,\rangle_{12}$ satisfies some desirable properties.

Let us first briefly recall the experimental set-ups and our theoretical point of view, the formalism, for the description of quantum teleportation and entanglement swapping.

\subsection{Quantum teleportation}\label{sec:quantum teleportation}

John Bell with his famous inequalities \cite{bell1964, bell-book} was the first to demonstrate the nonlocal feature of quantum mechanics. Mathematically, it is the entanglement of quantum states determined by analogous inequalities, the entanglement witness inequalities \cite{horodecki96, terhal00, bertlmann-narnhofer-thirring02, bruss02}, which provides the basis of quantum information processing, particularly in processes like quantum teleportation.

Quantum teleportation \cite{BBCJPW} together with its experimental verification \cite{bouwmeester-pan-zeilinger-etal, ursin-zeilinger-etal} is an amazing quantum feature that relies on the fact that in all finite dimensions several qudits can be entangled in different ways. Usually three two-dimensional qubits are considered together with the associated Bell states.

In order to demystify the ``hocus pocus'' \footnote{``Hocus pocus'' or more precise ``hocus pocus fidibus'' is a term originating from ancient Latin used in religious ceremonies and is nowadays spoken by magicians as magic formula to bring about some sort of change.} of the phenomenon, we first want to fix the rules.
The quantum states, described by vectors, are elements of a tensor product of three Hilbert spaces $\Ha_1 \otimes \Ha_2 \otimes \Ha_3$ of equal dimensions, the corresponding matrix algebras are denoted by $\A_1 \otimes \A_2 \otimes \A_3\,$. In the popular terminology, $\A_1, \A_2$ belong to Alice and $\A_3$ to Bob. Suppose that Alice gets in her first channel $\Ha_1$ an incoming message given by a vector $|\phi \rangle_1$ that she wants to transfer to Bob without direct contact between the algebras $\A_1$ and $\A_3$, but she knows how the vectors correspond to each other by a given isometry. To achieve this goal, she uses the fact that the three algebras can be entangled in different ways. Alice also knows that her second channel $\Ha_2$ is entangled with Bob by a source of entangled photons (called EPR source with reference to the work of Einstein, Podolsky and Rosen \cite{EPR}), such that the total state restricted to $\Ha_2 \otimes \Ha_3$ is maximally entangled.\\

\noindent\emph{Isometry:}

A maximally entangled state $|\psi \rangle_{23} \in \Ha_2 \otimes \Ha_3$ defines a map being an antilinear isometry $\tilde I_{32}$ between the vectors from one factor to the other \cite{Uhlmann01, TBKN}. More precisely, the antilinear isometry $\tilde I_{32}$ is a bijective map from an orthogonal basis $\{\,|\,\varphi_{i} \,\rangle_2 \,\}$ in $\Ha_2$ into a corresponding orthogonal basis $\{\,|\,\kappa_{i} \,\rangle_3 \,\}$ in $\Ha_3\,$. In our notation the first index of $\tilde I_{32}$ corresponds to the range of the map and the second one to the domain. The map is such that the components of the entangled state
\begin{equation}\label{state in H23}
|\,\psi\,\rangle_{23} \;=\; \frac{1}{\sqrt{d}}\,\sum\limits_{i=1}^d\,|\,\varphi_{i} \,\rangle_2 \otimes |\,\kappa_{i} \,\rangle_3
\end{equation}
are related by
\begin{equation}\label{isometry32}
|\,\tilde I_{32}\,\varphi_{i\,2} \,\rangle_3 \;=\; |\,\kappa_{i} \,\rangle_3 \quad \mbox{and} \quad |\,\tilde I_{32}\,U_2\,\varphi_{i\,2} \,\rangle_3 \;=\; |\,U_3^{\ast}\,\tilde I_{32}\,\varphi_{i\,2} \,\rangle_3 \;=\; U^{\ast}_3\,|\,\kappa_{i} \,\rangle_3\,,
\end{equation}
for every unitary operator $U\,$. Notice our choice of notation for the transpose, adjoint and complex conjugate of an operator: $(U_{ij})^{\top} = U_{ji}\,,\, (U_{ij})^{\dagger} =  (U_{ji})^{\ast}$ and $(U_{ij})^{\ast} = U_{ij}^{\ast}\,$.

Of course, the formalism also allows a reversed definition, which certainly has no impact on the physics. Then the antilinear isometry $\tilde I_{23}$ defines the map from the basis $\{\,|\,\kappa_{i} \,\rangle_3 \,\}$ in $\Ha_3$ into the basis $\{\,|\,\varphi_{i} \,\rangle_2 \,\}$ in $\Ha_2\,$, where the components are related by
\begin{equation}\label{isometry23}
|\,\tilde I_{23}\,\kappa_{i\,3} \,\rangle_2 \;=\; |\,\varphi_{i} \,\rangle_2 \quad \mbox{and} \quad |\,\tilde I_{23}\,U_3\,\kappa_{i\,3} \,\rangle_2 \;=\; |\,U_2^{\ast}\,\tilde I_{23}\,\kappa_{i\,3} \,\rangle_2 \;=\;U^{\ast}_2\,|\,\varphi_{i} \,\rangle_2\,.
\end{equation}
Clearly, isometry (\ref{isometry23}) is the inverse map of (\ref{isometry32}): $\tilde I_{23} \,=\, (\tilde I_{32})^{-1}\,$.\\

This metamorphosis of a vector first into an operator and then into an isomorphism (an isometry in our case) can be understood by a partial scalar product. The scalar product $\langle \,|\, \rangle$ occurs in the multiplication of the tensor product. If, however, the two vectors have a different number of factors (e.g. in the decomposition of basis vectors) there will remain a contribution from the scalar product $\langle \,\chi\, |_2 \,\big( \,|\,\varphi \,\rangle_2 \otimes |\,\kappa \,\rangle_3 \,\big) \,=\, \left\langle \,\chi\,|\,\varphi \,\right\rangle_2 \,|\,\kappa \,\rangle_3\,$. In the above case of decomposition (\ref{state in H23}), when considering the partial scalar product we obtain
\begin{equation}\label{transporter from H2 to H3}
\langle\,\chi\,|_2 \,\big( \,|\,\psi\,\rangle_{23} \,\big) \;=\; \frac{1}{\sqrt{d}}\, \sum\limits_{i=1}^d \,\left\langle \,\chi\,|\,\varphi_{i} \,\right\rangle_2 |\,\kappa_{i} \,\rangle_3 \;=\; |\,\phi\,\rangle_3 \,.
\end{equation}
In this way  a maximally entangled state $|\,\psi\,\rangle_{23}$ becomes a map, a \emph{quantum transporter} from $\Ha_2 \rightarrow \Ha_3\,$.\\

The possibility to transfer the incoming state of $\Ha_1$ at Alice into a state of $\Ha_3$ at Bob uses the fact that an isometry $I_{31}$ between these two algebras was taken for granted. Now Alice chooses an isometry $\tilde I_{21}\,$, which corresponds to choosing a maximally entangled state $|\psi \rangle_{12} \in \Ha_1 \otimes \Ha_2\,$, such that the following isometry relation holds
\begin{equation}\label{isometry relation for 1-3}
\tilde I_{32} \circ \tilde I_{21} \;=\; I_{31}\,.
\end{equation}
We denote the antilinear isometry by tilde, the linear one without, and the composition of two maps by a circle $\circ \,$. The operation $\circ \,$ means the map $\tilde I_{21}$ is followed by $\tilde I_{32}\,$. Note, the operation $\circ \,$ is a pure mathematical composition of maps and need not correspond to the time-ordering of the physical processes, which is given by the order of successive projection operators.

\begin{figure}
	\centering
	\includegraphics[width=0.4\textwidth]{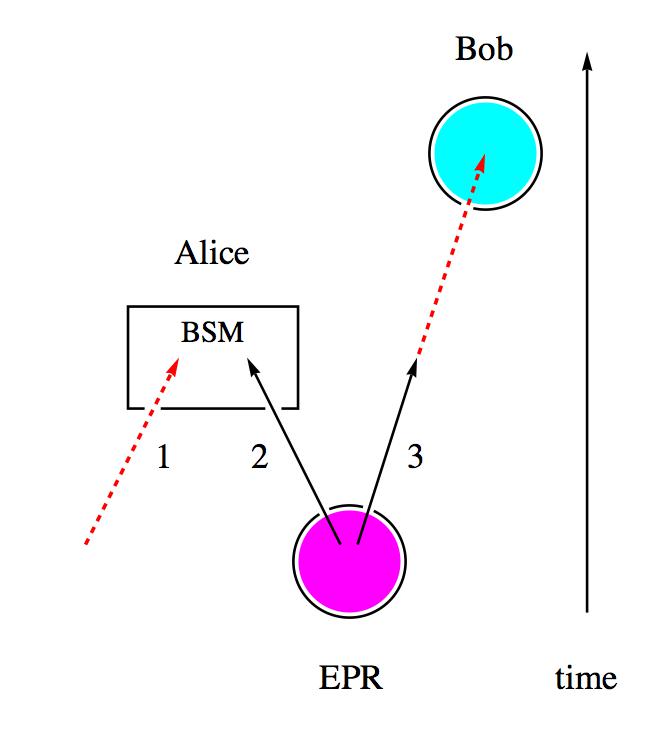}
	\caption{Quantum teleportation. A pair of entangled photons is emitted into the channels (2,3) by an EPR source. Independently, a photon in an arbitrary state in channel 1 arrives at Alice's side together with the photon in channel 2 of the EPR pair. When Alice performs a Bell state measurement in channels (1,2) the state of photon 1 is instantaneously teleported to photon 3 at Bob's side (sketched by broken arrows) . However, which Bell state Alice is measuring (there are four possibilities) she has to communicate to Bob in a classical way so that Bob can perform the appropriate unitary transformation (adjust his apparatus) to find his photon in the state of the incoming one.}
	\label{fig:quantum teleportation}
\end{figure}

Expressed in $\{ \varphi_i \}\,$, an orthonormal basis (ONB) of one factor, the state vector can be written as
\begin{equation}\label{psi-12 isometry}
|\,\psi\,\rangle_{12} \;=\; \frac{1}{\sqrt{d}}\,\sum\limits_{i=1}^d \,|\,\varphi_{i} \,\rangle_1 \otimes | \,\tilde I_{21} \,\varphi_{i\,1} \,\rangle_2 \;.
\end{equation}
A measurement by Alice in $\Ha_1 \otimes \Ha_2$ with the outcome of entangled state (\ref{psi-12 isometry}) produces the desired state in $\Ha_3\,$, i.e., the incoming state $|\,\phi\,\rangle_1 \in \Ha_1$ at Alice has been teleported to Bob: $|\,\phi\,\rangle_3 \in \Ha_3\,$, see Fig.~\ref{fig:quantum teleportation}$\,$. Mathematically, it is described by the linear isometry $I_{31}$ (\ref{isometry relation for 1-3})$\,$.

The outcome of other maximally entangled states, orthogonal to the first one, corresponds to a unitary transformation of the form $U_{12}= U_1 \otimes \mathds{1}_2$ in $\Ha_1 \otimes \Ha_2\,$, which produces a unique unitary transformation $U_{3}\,$ in $\Ha_3$ that Bob can perform to obtain the desired state. Thus Alice just has to tell Bob her measurement outcome via some classical channel. The measurements of Alice produce the following results for Bob:
\begin{eqnarray}\label{eq:measurements of Alice in teleportation}
\big(\left|\,\psi\,\right\rangle\left\langle\,\psi\,\right|\big)_{12} \otimes \mathds{1}_{3} \;|\,\phi\,\rangle_1 \otimes |\,\psi\,\rangle_{23} &\;=\;& \frac{1}{d}\,|\,\psi\,\rangle_{12} \otimes |\,\phi\,\rangle_{3} \,,\\
U_{12}\,\big(\left|\,\psi\,\right\rangle\left\langle\,\psi\,\right|\big)_{12}\,U_{12}^{\dagger}\, \otimes \mathds{1}_{3} \;|\,\phi\,\rangle_1 \otimes |\,\psi\,\rangle_{23} &\;=\;& \frac{1}{d}\,U_{12}|\,\psi\,\rangle_{12} \otimes U_{3}|\,\phi\,\rangle_{3} \;,
\end{eqnarray}
with the unitary transformation $U_{12}= U_1 \otimes \mathds{1}_2$ in $\Ha_1 \otimes \Ha_2\,$.

The state vectors can be expressed by an ONB for a fixed isometry, e.g., choosing $\tilde I_{32}$ in the following way:
\begin{eqnarray}\label{state vectors in teleportation}
|\,\psi\,\rangle_{23} &\;=\;& \frac{1}{\sqrt{d}}\,\sum\limits_{i=1}^d\,|\,\varphi_{i} \,\rangle_2 \otimes |\,\varphi_{i} \,\rangle_3 \;,\\
|\,\phi\,\rangle_{1 \,\rm{or}\, 3} &\;=\;& \sum\limits_{i=1}^d\,\alpha_{i}\,|\,\varphi_{i} \,\rangle_{1 \,\rm{or}\, 3} \;.\label{state vector teleported}
\end{eqnarray}
The antilinear isometries $\tilde I_{32}$ and $\tilde I_{21}$ corresponding to the two given maximally entangled states combine to the linear isometry $I_{31} \;=\; \tilde I_{32} \circ \tilde I_{21}$ defining the map in Eq.~(\ref{state vector teleported}). Changing to another Bell state in Alice's measurement corresponds to
\begin{equation}\label{isometry relation for other Bell states}
\tilde I_{32} \circ \tilde I_{21} \,U_1 \;=\; \tilde I_{32} \circ U_2^{\ast}\,\tilde I_{21} \;=\; U_3 \,\tilde I_{32} \circ \tilde I_{21}
\;=\; U_3 \, I_{31} \;.
\end{equation}

Summarizing, if Alice measures the same Bell state in the sense of relation (\ref{isometry relation for 1-3}) between $\Ha_1$ and $\Ha_2$ as there was between $\Ha_2$ and $\Ha_3$, which was given by the EPR source, she knows that her measurement left Bob's $\Ha_3$ in the state $|\phi \rangle_3\,$ which is Alice's incoming state. If Alice finds, on the other hand, a different Bell state, which is given by $U_{12}|\psi \rangle_{12}$ (and $U_{12}= U_1 \otimes \mathds{1}_2$) since all other Bell states are connected by unitary transformations, then Bob will have the state vector $U_{3}|\phi \rangle_{3}$, where the unitary transformation $U_{3}$ is determined by $U_{12}\,$.

\subsection{Entanglement swapping}\label{sec:entanglement swapping}

\begin{figure}
	\centering
	\includegraphics[width=0.45\textwidth]{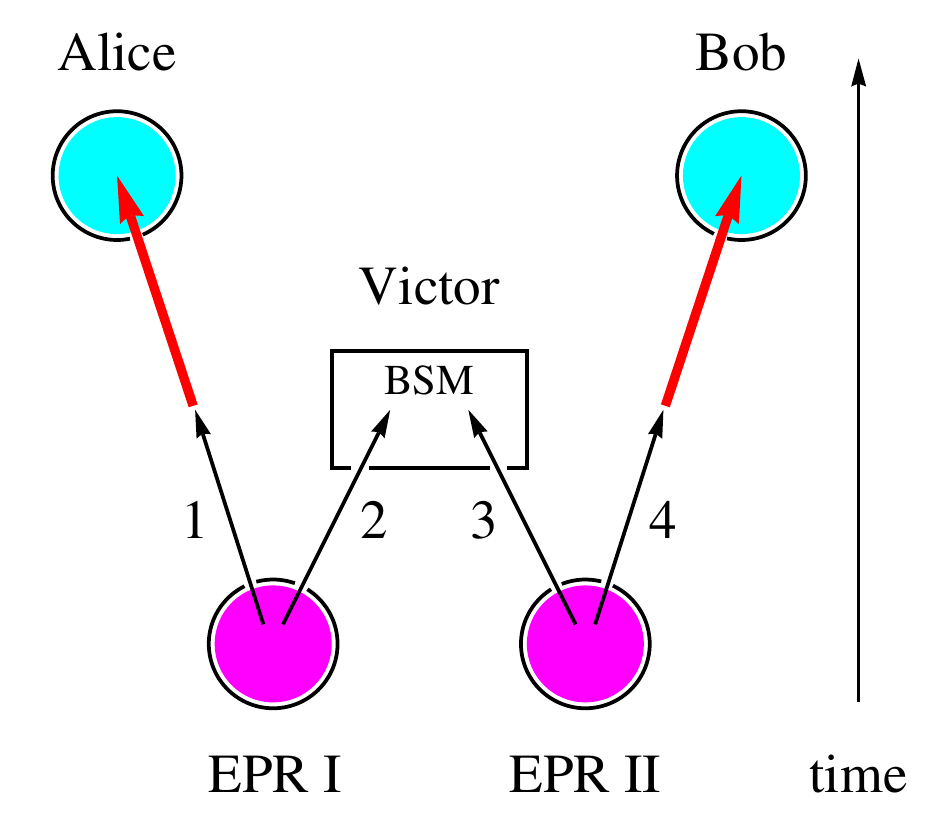}
	\caption{Entanglement swapping. Two pairs of entangled photons are emitted by the sources EPR~I and EPR~II$\,$. When Victor entangles two photons, i.e. performs a Bell state measurement in channels (2,3), the remaining two photons in channels (1,4) become instantaneously entangled into the same entangled state -- the entanglement swapped -- which is measured by Alice and Bob.}
	\label{fig:entanglement swapping}
\end{figure}

Closely related to teleportation of single quantum states is another striking quantum phenomenon called entanglement swapping \cite{zukowski-zeilinger-horne-ekert}. It can be interpreted as teleportation of an unspecified state (without well-defined polarization properties) in channel 2 onto the photon in channel 4, or of the unspecified state in channel 3 onto the photon in channel 1, see Fig.~\ref{fig:entanglement swapping}. Altogether, the entanglement of the photons created by a Bell state measurement in channels (2,3) swaps onto the photons in channels (1,4). Experimentally, entanglement swapping has been demonstrated in Ref.~\cite{pan-bouwmeester-weinfurter-zeilinger} and is nowadays a standard tool in quantum information processing \cite{bertlmann-zeilinger02}. The set-up is described by a tensor product of four Hilbert spaces $\Ha_1 \otimes \Ha_2 \otimes \Ha_3 \otimes \Ha_4$ of equal dimensions. The entanglement swapping phenomenon illustrates the different slicing possibilities of a 4-fold tensor product into $(1,2) \otimes (3,4)$ or $(1,4) \otimes (2,3)\,$, where, e.g., $(1,2)$ denotes entanglement between the subsystems $\Ha_1$ and $\Ha_2\,$, see Fig.~\ref{fig:entanglement swapping}.

The same way of reasoning as in Sec.~\ref{sec:quantum teleportation} can be applied to entanglement swapping, i.e., we consider a maximally entangled state as an isometry between the vectors of one factor to the other. In case of entanglement swapping the starting point are two maximally entangled pure states combined in a tensor product $|\,\psi\,\rangle_{12} \otimes |\,\psi\,\rangle_{34}\,$. Expressed in an ONB of one factor, the entangled states are given by Eq.~(\ref{psi-12 isometry}). They describe usually two pairs of EPR photons. These propagate into different directions and at the interaction point of two of them, say photon $2$ and $3$, Victor performs a Bell state measurement (BSM), i.e., a measurement with respect to an orthogonal set of maximally entangled states. Usually they are chosen as the Bell states $\left|\,\psi^{\,\pm}\,\right\rangle_{23} \,=\, \frac{1}{\sqrt{2}}\,\left(\,
		\left|\,\uparrow\,\right\rangle_2\otimes\left|\,\downarrow\,\right\rangle_3\,\pm\,
		\left|\,\downarrow\,\right\rangle_2\otimes\left|\,\uparrow\,\right\rangle_3\,\right)$
and $\left|\,\phi^{\,\pm}\,\right\rangle_{23} \,=\, \frac{1}{\sqrt{2}}\,\left(\,
		\left|\,\uparrow\,\right\rangle_2\otimes\left|\,\uparrow\,\right\rangle_3\,\pm\,
		\left|\,\downarrow\,\right\rangle_2\otimes\left|\,\downarrow\,\right\rangle_3\,\right)\,$.
But our results are more general, the four factors just have to be of the same dimension $d\,$, which is arbitrary.

In complete analogy to the case of teleportation, discussed before, the effect of the projection corresponding to the measurement on the state is
\begin{equation}\label{eq:measurement for 2-3 in entanglement swapping}
\big(\left|\,\psi\,\right\rangle\left\langle\,\psi\,\right|\big)_{23} \otimes \mathds{1}_{14} \;|\,\psi\,\rangle_{12} \otimes |\,\psi\,\rangle_{34} \;=\; \frac{1}{d}\,|\,\psi\,\rangle_{23} \otimes |\,\psi\,\rangle_{14} \;,
\end{equation}
where $|\psi \rangle_{14} \in \Ha_1 \otimes \Ha_4\,$ is a maximally entangled state that can be expressed in an ONB $\{ \varphi_i \}\,$ of one factor
\begin{equation}\label{psi-14 isometry}
|\,\psi\,\rangle_{14} \;=\; \frac{1}{\sqrt{d}}\,\sum\limits_{i=1}^d \,|\,\varphi_{i} \,\rangle_1 \otimes | \,\tilde I_{41} \,\varphi_{i\,1} \,\rangle_4 \;.
\end{equation}
It corresponds to the antilinear isometry $\tilde I_{41}$ satisfying the relation
\begin{equation}\label{isometry relation for 1-4}
\tilde I_{41} \;=\; \tilde I_{43} \circ \tilde I_{32} \circ \tilde I_{21} \;.
\end{equation}
The other isometries $\tilde I_{43}, \tilde I_{32}, \tilde I_{21}\,$ correspond to the other maximally entangled states.

Thus, after the Bell state measurement of photon $2$ and $3$ into a definite entangled state the photons $1$ and $4$ become instantaneously entangled into the same state, see Fig.~\ref{fig:entanglement swapping}$\,$. Remarkably, the two photons originate from different noninteracting sources.

\section{Delayed-Choice of Entanglement Swapping}\label{sec:delayed choice of entanglement swapping}

Now we are prepared to turn to the phenomenon of delayed-choice entanglement swapping \cite{Ma-Zeilinger,Peres00}, which is interpreted as being quite paradoxical. In this experiment the order of the measurements is reversed as compared to entanglement swapping. Alice and Bob measure first and record their data, then at a later time Victor is free to choose a projection of his two states onto an entangled state or to measure them individually. The outcome of the measurements of Alice, Bob and Victor is recorded and compared to the previous case where Victor measured before Alice and Bob. It turns out that the joint probability for the outcome of these two cases is the same. Particularly in case of delayed-choice, it is Victor's measurement that decides the context and determines the interpretation of Alice's and Bob's data. Alice and Bob can sort their already recorded data, according to Victor's later choice and his results, in such a way that they can verify either the entangled or the separable states.

\subsection{General Discussion}\label{sec:general discussion}

To analyze this delayed-choice procedure, we prefer to use the density matrix formalism for the description of the quantum states. We start with a tensor product of four matrix algebras $\A_{\rm{tot}} = \A_1 \otimes \A_2 \otimes \A_3 \otimes \A_4$ of equal dimensions; $\A_1$ and $\A_4$ belong to Alice and Bob and the subalgebra $\A_2 \otimes \A_3$ refers to Victor's algebra.

At the beginning the state described by the density matrix $\rho \in \A_{\rm{tot}}$ is separable on $\A_1 \otimes \A_4$ and by manipulation  -- Bell state measurement -- in $\A_2 \otimes \A_3$ the state over $\A_1 \otimes \A_4$ becomes entangled.

The measurements correspond to the projection operators
$Q^{\,i}_{23} \,=\, \big(\left|\,\psi^i\,\right\rangle\left\langle\,\psi^i\,\right|\big)_{23}$
(with i = 1, \dots ,4) onto the maximally entangled Bell states
$|\psi^{1,2} \rangle_{23} = \left|\,\psi^{\,\pm}\,\right\rangle_{23}$ and
$|\psi^{3,4} \rangle_{23} = \left|\,\phi^{\,\pm}\,\right\rangle_{23}\,$.
Thus the operators $Q^{\,i}_{23}$ form a Bell basis in subalgebra $\A_2 \otimes \A_3\,$.

If we project on such a basis we immediately get a new state on $\A_1 \otimes \A_4\,$, i.e., we have a collapse of the quantum state, which is up to normalization given by
\begin{equation}\label{rho projected by Q}
Q^{\,i}_{23}\,\rho\,Q^{\,i}_{23} \,.
\end{equation}

Let us start with the case where $\rho \in \A_{\rm{tot}}$ is separable on $\A_1 \otimes \A_2$ and maximally entangled on $\A_3 \otimes \A_4\,$,
\begin{equation}\label{rho-sep12-entangle34 project to Q23}
\rho \;=\; \rho_1\otimes\rho_2\otimes\rho_{34} \;\;\stackrel{Q^{\,i}_{23}\;}{\longrightarrow}\;\; \rho_1\otimes\rho^{\,i}_{23}\otimes\rho^{\,i}_{4} \,,
\end{equation}
where $\rho_{34} = \big(\left|\,\psi\,\right\rangle\left\langle\,\psi\,\right|\big)_{34}$ corresponds to some entangled state that at its best can be a Bell state on the algebra $\A_3 \otimes \A_4$ and $\rho^{\,i}_{23} \equiv Q^{\,i}_{23}$. We find $\rho_1\otimes\rho^{\,i}_{4}$ separable with a corresponding probability depending on the projector $Q^{\,i}_{23}$ and on the initial state $\rho_{34}$.

Next we consider the case where $\rho \in \A_{\rm{tot}}$ is maximally entangled on $\A_1 \otimes \A_2$ \emph{and} on $\A_3 \otimes \A_4\,$. If we perform a BSM on $\A_2 \otimes \A_3$ then we have
\begin{equation}\label{rho-entangled12-entangled34 project to Q23}
\rho \;=\; \rho_{12}\otimes\rho_{34} \;\;\stackrel{Q^{\,i}_{23}\;}{\longrightarrow}\;\; Q^{\,i}_{23}\,\rho\,Q^{\,i}_{23} \;=\; \rho^{\,i}_{14}\otimes\rho^{\,i}_{23} \,.
\end{equation}
We now find the state $\rho^{\,i}_{14}$ maximally entangled depending on the projector $Q^{\,i}_{23}$ and corresponding to the antilinear isometry $\tilde I^{\,i}_{41}\,$, which satisfies -- as before in Eq.~(\ref{isometry relation for 1-4}) -- the relation
\begin{equation}\label{isometry relation for 1-4 with i}
\tilde I^{\,i}_{41} \;=\; \tilde I_{43} \circ \tilde I^{\,i}_{32} \circ \tilde I_{21} \;.
\end{equation}
Thus, after the projection the state $\big(Q^{\,i}_{23}\,\rho\,Q^{\,i}_{23}\big)_{\A_1 \otimes \A_4}$ is pure and maximally entangled on $\A_1 \otimes \A_4\,$; the entanglement has swapped.\\

It gives us the idea that with help of entanglement swapping we should be able to check to what extent the collapse of a quantum state happens. Let us consider a sequence of measurements where the results emerge with a certain probability.\\

\noindent\emph{Question:}

To what extent does the probability of the measurement outcome depend on the chronological order of the measurements?

\noindent\emph{Classical physics:}

Consider the time-invariant states in classical physics. There clearly the measurements are independent of the chosen time, thus independent of the order of the successive measurements.

\noindent\emph{Quantum physics:}

In quantum physics, however, there exists a \emph{quantum history} \cite{Gell-Mann_Hartle, Griffiths}. Consider a quantum state on some larger algebra and several measurements on that state. The measurements we describe by projectors
$P^{\,i}_1 = \big(\left|\,\psi^i\,\right\rangle\left\langle\,\psi^i\,\right|\big)_{1}$ and $Q^{\,j}_2 = \big(\left|\,\phi^j\,\right\rangle\left\langle\,\phi^j\,\right|\big)_2$
on some states of the subalgebras $\A_1, \A_2\,$. We perform the measurements $M_1 = \{P^{\,i}_1\}$ and $M_2 = \{Q^{\,j}_2\}$
and obtain
\begin{equation}\label{rho measurement by P_1 and Q_2}
\rho \stackrel{M_1\;}{\longrightarrow} \;\; \rho^{\,i} \;=\; P^{\,i}_1\,\rho\,P^{\,i}_1 \;\; \stackrel{M_2\;}{\longrightarrow} \;\; \rho^{\,ji} \;=\; Q^{\,j}_2 P^{\,i}_1\,\rho\,P^{\,i}_1 Q^{\,j}_2 \,.
\end{equation}
The state $\rho^{\,i}$ after the first measurement $M_1$ occurs with a probability $w^i = {\rm{Tr}}\,P^{\,i}_1\,\rho\,P^{\,i}_1$ and the state $\rho^{\,ji}$ after the second measurement $M_2$ with a probability $w^{\,ji} = {\rm{Tr}}\,Q^{\,j}_2 P^{\,i}_1\,\rho\,P^{\,i}_1 Q^{\,j}_2 = {\rm{Tr}}\,\rho\,P^{\,i}_1 Q^{\,j}_2 P^{\,i}_1\,$.

Now we interchange the order of measurements $M_1 \longleftrightarrow M_2\,$. Then the state
\begin{equation}\label{state rho ij}
\rho^{\,ij} \;=\; P^{\,i}_1 Q^{\,j}_2 \,\rho\, Q^{\,j}_2 P^{\,i}_1
\end{equation}
occurs with probability $w^{\,ij} = {\rm{Tr}}\,\rho\, Q^{\,j}_2 P^{\,i}_1 Q^{\,j}_2\,$. In case of commuting projection operators there is no dependence on the order of measurement operations, i.e.,
\begin{equation}\label{commutator P1Q2 vanishing equal probabilities}
\left[\,P^i_1\,,\,Q^j_2\,\right] \;=\; 0 \qquad \Longrightarrow \qquad w^{\,ij} = w^{\,ji} \,.
\end{equation}
The probabilities of the measurement outcomes, which can be determined from the protocols of the measurements, are the same! Actually, it is enough to examine whether the condition ${\rm{Tr}}\,\rho\, \big(Q^{\,j}_2 P^{\,i}_1 Q^{\,j}_2 - P^{\,i}_1 Q^{\,j}_2 P^{\,i}_1\big) = 0$ holds.

\subsection{Experiment of Zeilinger's Group}\label{sec:experiment Zeilinger group}

Now we turn to the experiment of the Zeilinger group \cite{Ma-Zeilinger}. Let us begin with the case of entanglement swapping. Thus initially we have the two Bell states $\rho_{12}$ and $\rho_{34}$ and first Victor performs a BSM  $M_1 = \{Q^{\,i}_{23}\}$ on some maximally entangled state $\rho^{\,i}_{23} \equiv Q^{\,i}_{23}$ on $\A_2 \otimes \A_3\,$. Then we obtain as above, Eq.~(\ref{rho-entangled12-entangled34 project to Q23}),
\begin{equation}\label{rho-entangled12-entangled34 first M1}
\rho \;\;\stackrel{M_1\;}{\longrightarrow}\;\; \rho^{\,i} \;=\; Q^{\,i}_{23}\,\rho_{12}\otimes\rho_{34}\,Q^{\,i}_{23} \;=\; \rho^{\,i}_{14}\otimes\rho^{\,i}_{23} \,,
\end{equation}
where the maximally entangled state $\rho^{\,i}_{14}$ on $\A_1 \otimes \A_4$ occurs with probability $w^i = \frac{1}{4}$ and satisfies the isometry relation (\ref{isometry relation for 1-4 with i}) between the algebras.

The second measurement $M_2 = \{P^{\,ab}_{14} = P^{\,a}_1 \otimes P^{\,b}_4\}$ is performed by Alice and Bob who measure the incoming states at $\A_1$ and $\A_4$ by projecting on some states $|\psi^{a} \rangle_1$ and $|\psi^{b} \rangle_4\,$. Denoting the corresponding projectors by $P^{\,a}_1 = \big(\left|\,\psi^a\,\right\rangle\left\langle\,\psi^a\,\right|\big)_{1}$ and $P^{\,b}_4 = \big(\left|\,\psi^b\,\right\rangle\left\langle\,\psi^b\,\right|\big)_{4}\,$, the state turns into (see Fig.~\ref{fig:entanglement swapping})
\begin{equation}\label{rho measurement by Pa_1 and Pb_4}
\rho^{\,i} \;\; \stackrel{M_2\;}{\longrightarrow} \;\; \rho^{\,iab} \;=\; P^{\,ab}_{14} Q^{\,i}_{23} \,\rho\, Q^{\,i}_{23} P^{\,ab}_{14} \;=\; \rho^{\,a}_1 \otimes \rho^{\,i}_{23} \otimes \rho^{\,b}_4 \,.
\end{equation}
The probability $w^{\,iab}$ of finding the state $\rho^{\,iab}$ is determined by the isometry $\tilde I^{\,i}_{41}\,$, i.e., it depends on the chosen maximally entangled state denoted by $i\,$.\\

\begin{figure}
	\centering
	\includegraphics[width=0.45\textwidth]{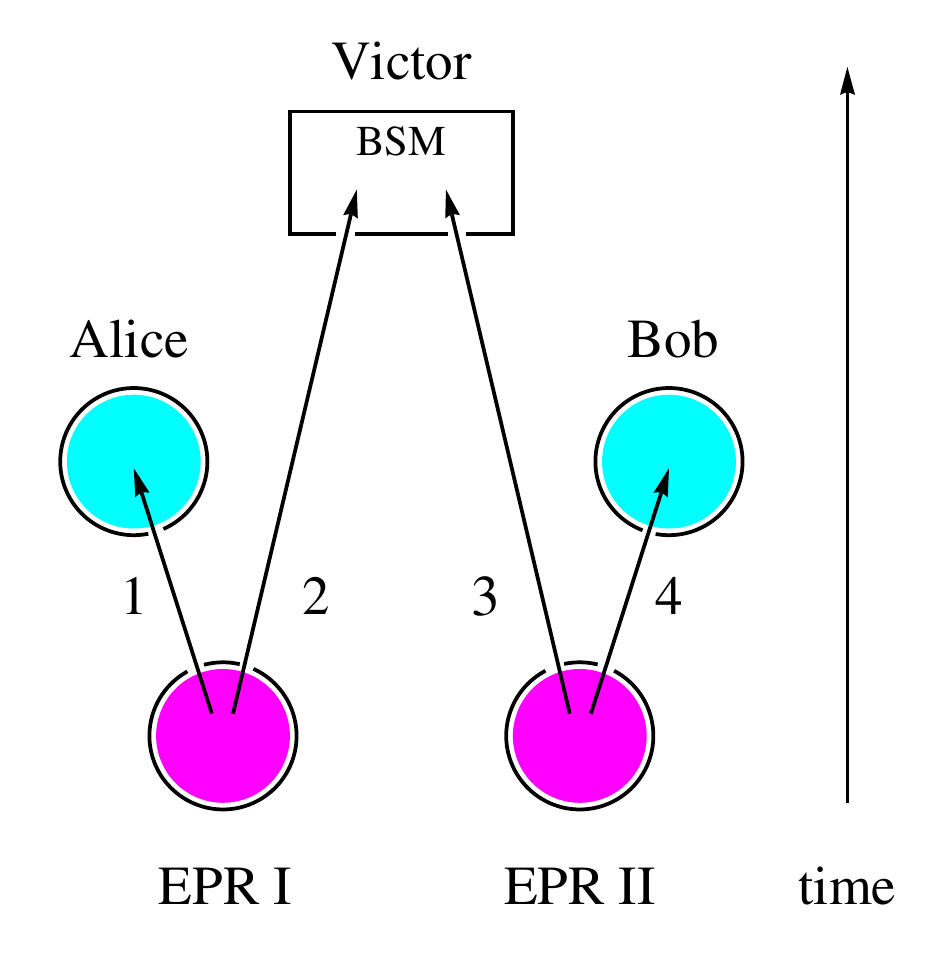}
	\caption{Delayed-choice entanglement swapping. There are two EPR sources I and II and we have a reversed order of measurements. Alice and Bob measure first in channels (1,4) and record their data and later on Victor projects in a free choice his state in channels (2,3) onto an entangled or a separable state. When comparing this case with entanglement swapping of Fig.~\ref{fig:entanglement swapping}$\,$, it turns out that the joint probability for the outcome of these two cases is the same due to the commutativity of the projection operators, i.e., there is independency of the order of successive measurements.}
	\label{fig:delayed entanglement swapping}
\end{figure}

Next we study the reversed order of measurements, i.e., the case of delayed-choice entanglement swapping (see Fig.~\ref{fig:delayed entanglement swapping}).

The first measurement is now $M_2 = \{P^{\,ab}_{14} = P^{\,a}_1 \otimes P^{\,b}_4\}$ performed by Alice and Bob. They measure their incoming states by projecting on some states $|\psi^{a} \rangle_1$ and $|\psi^{b} \rangle_4\,$, then the total state $\rho$ turns into
\begin{equation}\label{rho-entangled12-entangled34 first M2}
\rho \;\;\stackrel{M_2\;}{\longrightarrow}\;\; \rho^{\,ab} \;=\; P^{\,ab}_{14}\,\rho_{12} \otimes \rho_{34}\,P^{\,ab}_{14} \;=\; \rho^{\,a}_1 \otimes \rho^{\,a}_2 \otimes \rho^{\,b}_3 \otimes \rho^{\,b}_4 \,,
\end{equation}
which is separable and even pure on the subalgebra $\A_1 \otimes \A_4\,$.

In a second measurement Victor may choose freely to project onto an entangled state or onto a separable state on $\A_2 \otimes \A_3\,$. Let us discuss first the measurement $M_1 = \{Q^{\,i}_{23}\}$ onto an entangled state. Then the above state changes into
\begin{equation}\label{rho-entangled12-entangled34 first M2 second M1}
\rho^{\,ab} \;\;\stackrel{M_1\;}{\longrightarrow}\;\; \rho^{\,abi} \;=\; Q^{\,i}_{23} P^{\,ab}_{14}\,\rho_{12} \otimes \rho_{34}\,P^{\,ab}_{14} Q^{\,i}_{23} \;=\; \rho^{\,a}_1 \otimes \rho^{\,i}_{23} \otimes \rho^{\,b}_4 \,,
\end{equation}
which remains separable on $\A_1 \otimes \A_4\,$. The entangled state $\rho^{\,i}_{23}$ clearly depends on the isometry $\tilde I^{\,i}_{32}(a,b)$ and on the chosen $(a,b)\,$.

Since the projection operators -- the measurement operators -- commute,
\begin{equation}\label{commutator P14Q23 vanishing equal probabilities}
\left[\,P^{\,ab}_{14}\,,\,Q^{\,i}_{23}\,\right] \;=\; 0 \qquad \Longrightarrow \qquad w^{\,abi} = w^{\,iab} \,,
\end{equation}
the probability $w^{\,abi} = {\rm{Tr}}\,\rho\,P^{\,ab}_{14} Q^{\,i}_{23} P^{\,ab}_{14}$ for the occurrence of the final state $\rho^{\,abi}$ is the same as before. In fact, $\rho^{\,abi} = \rho^{\,iab}$ and the results are independent of the order of the measurements. It is valid for all projection operators $P^{\,ab}_{14}$ independently of their chosen orientations $a,b\,$. Therefore, different quantum histories lead to the same result.

In case Victor projects onto a separable state on $\A_2 \otimes \A_3\,$, we have as measurement operators ${\widehat{M}}_1 = \{P^{\,j}_{23}\}\,$, where the projectors
$P^{\,j}_{23} = \big(\left|\,\phi^j\,\right\rangle\left\langle\,\phi^j\,\right|\big)_{23}$
project onto the separable states
$\left|\,\phi^{\,1,2,3,4}\,\right\rangle_{23} = \{
		\left|\,\uparrow\,\right\rangle_2\otimes\left|\,\uparrow\,\right\rangle_3 \,,
		\left|\,\downarrow\,\right\rangle_2\otimes\left|\,\downarrow\,\right\rangle_3 \,, \left|\,\uparrow\,\right\rangle_2\otimes\left|\,\downarrow\,\right\rangle_3\,,
		\left|\,\downarrow\,\right\rangle_2\otimes\left|\,\uparrow\,\right\rangle_3 \}\,$.
Then again the commutator of the projectors is commuting,
\begin{equation}\label{commutator P14P23 vanishing equal probabilities}
\left[\,P^{\,ab}_{14}\,,\,P^{\,j}_{23}\,\right] \;=\; 0 \qquad \Longrightarrow \qquad w^{\,abj} = w^{\,jab} \,,
\end{equation}
and again we find independency of the order of measurements.

\section{Teleportation and Delayed Entanglement Swapping}\label{sec:teleportation and delayed entanglement swapping}

In this section we are going to use entanglement swapping as a source for quantum teleportation between Alice and Bob. As we shall see, it is not so much about entanglement swapping -- delayed or not -- but rather about double teleportation. We will find two cases, one where the order of successive measurements does not matter, and another one where it does corresponding to noncommutative projection operators.

We consider here a tensor product of five matrix algebras $\A_{\rm{tot}} = \A_0 \otimes \A_1 \otimes \A_2 \otimes \A_3 \otimes \A_4$ of equal dimensions, where we have an additional (general) incoming state
\begin{equation}\label{phi0 at A0}
\rho_0 \;=\; \big(\left|\,\phi\,\right\rangle\left\langle\,\phi\,\right|\big)_0
\end{equation}
on the algebra $\A_0$ at Alice's side, which we can write as
\begin{equation}\label{vector phi0 at A0}
\left|\,\phi\,\right\rangle_0 \,=\, \,\alpha \left|\,\uparrow\,\right\rangle_0 \,+\, \beta \left|\,\downarrow\,\right\rangle_0
\quad \mbox{and} \quad |\alpha|^2 + |\beta|^2 = 1\,,
\end{equation}
if we are dealing with qubits. If now Alice and Victor perform a BSM the incoming state $\left|\,\phi\,\right\rangle_0$ at Alice's side is teleported into $|\,\phi\,\rangle_4 \,=\, \left|\,I_{40}\,\phi_0\,\right\rangle_4$ at Bob's side.

\subsection{Commutative Projection Operators}\label{commutative projection operators}

\begin{figure}
	\centering
	\includegraphics[width=0.50\textwidth]{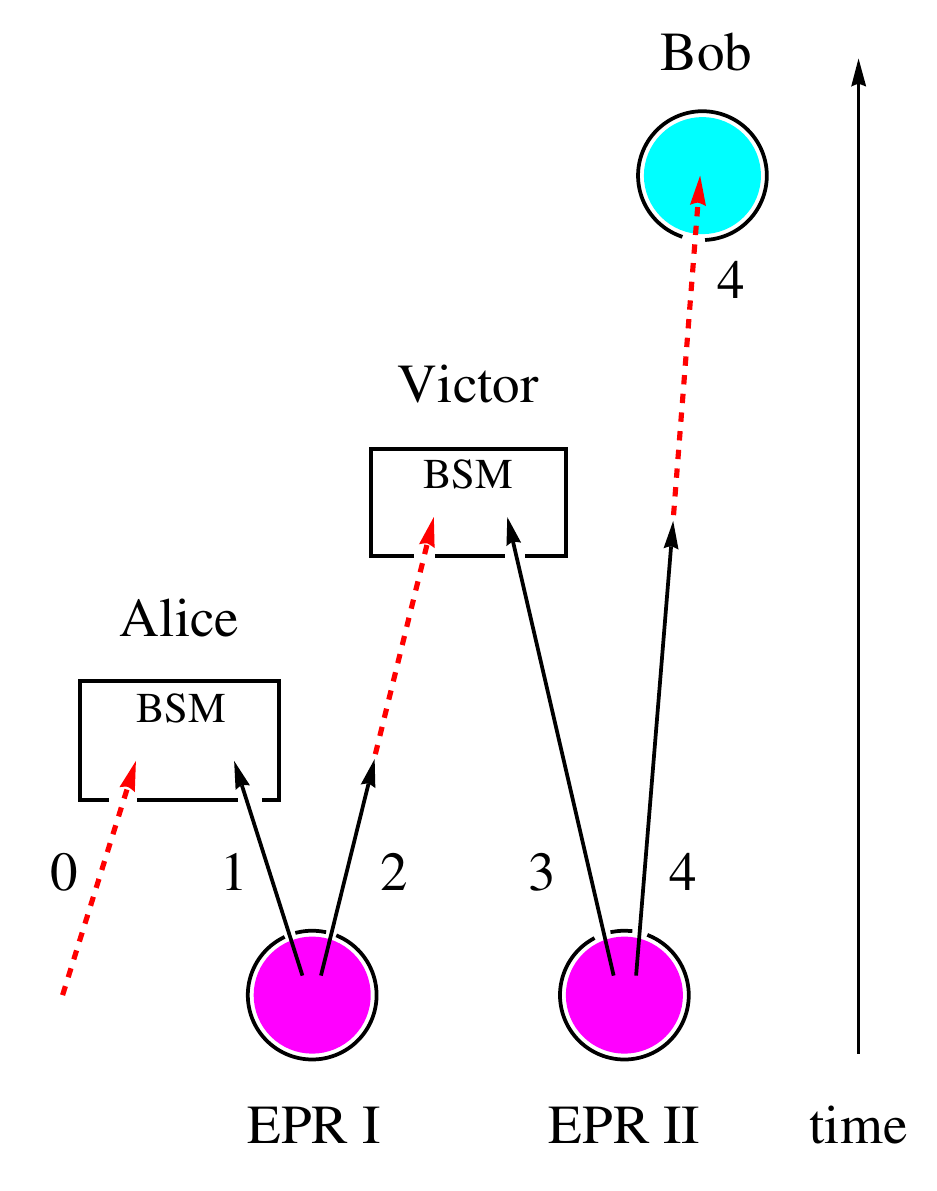}
	\caption{Teleportation with delayed entanglement swapping. There are two EPR sources I and II and we have double quantum teleportation of an incoming state at channel 0 to the same outgoing state at channel 4, which is measured by Bob. In between there is first a Bell state measurement by Alice in the channels (0,1), teleporting the incoming state from 0 to 2, and second a measurement by Victor in channels (2,3), producing a kind of delayed entanglement swapping, which teleports the state from 2 to 4$\,$.}
	\label{fig:teleportation with delayed entanglement swapping}
\end{figure}

We consider again a physical situation similar to Sec.~\ref{sec:experiment Zeilinger group} -- a kind of delayed entanglement swapping -- where we have two EPR sources providing the Bell states $\rho_{12}$ and $\rho_{34}\,$, but now an additional particle in the quantum state (\ref{phi0 at A0}) enters on Alice's side, see Fig.~\ref{fig:teleportation with delayed entanglement swapping}$\,$.

Firstly, Alice performs a BSM, $M_{\rm{Alice}} = \{Q_{01}\}$ with the corresponding projection operator $Q_{01} \,=\, \big(\left|\,\psi\,\right\rangle\left\langle\,\psi\,\right|\big)_{01}\,$. She projects onto some maximally entangled state $\rho_{01} \equiv Q_{01}$ on $\A_0 \otimes \A_1\,$. Then the total state being initially
\begin{equation}\label{rho total initially}
\rho_{\rm{initial}} \;=\; \rho_0 \otimes \rho_{12} \otimes \rho_{34}
\end{equation}
changes into
\begin{equation}\label{rho-add0-entangled12-entangled34 first M-Alice}
\rho_{\rm{initial}} \;\;\stackrel{M_{\rm{Alice}}\quad\;}{\longrightarrow}\;\;
Q_{01} \,\rho_0 \otimes \rho_{12} \otimes \rho_{34} \, Q_{01} \;=\; \rho_{01} \otimes \rho_2 \otimes \rho_{34} \,,
\end{equation}
where the state
\begin{eqnarray}\label{rho-phi-2}
\rho_2 &\;=\;& \big(|\,\phi\,\rangle\langle\,\phi\,|\big)_2 \;=\;
\big(\left|\,I_{20}\,\phi_0\,\right\rangle\left\langle\,I_{20}\,\phi_0\,\right|\big)_2
\end{eqnarray}
satisfies the isometry relation
\begin{equation}\label{isometry relation I02}
I_{20} \;=\; \tilde I_{21} \circ \tilde I_{10} \,.
\end{equation}
Thus the incoming state $\rho_0$ (\ref{phi0 at A0}) is teleported isometrically from $\A_0$ to $\A_2\,$, see Fig.~\ref{fig:teleportation with delayed entanglement swapping}.

In a second measurement $M_{\rm{Victor}} = \{Q_{23}\}$ with the corresponding projection operator $Q_{23} \,=\, \big(\left|\,\psi\,\right\rangle\left\langle\,\psi\,\right|\big)_{23}$ (again a BSM), Victor chooses freely to project onto some maximally entangled state $\rho_{23} \equiv Q_{23}$ on $\A_2 \otimes \A_3\,$. Then the state (\ref{rho-add0-entangled12-entangled34 first M-Alice}) turns into
\begin{equation}\label{rho-add0-entangled12-entangled34 first M-Alice second M-Victor}
\rho \;=\; \rho_{01} \otimes \rho_2 \otimes \rho_{34} \;\;\stackrel{M_{\rm{Victor}}\quad\;\;}{\longrightarrow}\;\;
Q_{23} \,\rho_{01} \otimes \rho_2 \otimes \rho_{34}\, Q_{23} \;=\; \rho_{01} \otimes \rho_{23} \otimes \rho_4 \,,
\end{equation}
where the state $\rho_4$ is given by
\begin{eqnarray}\label{rho-phi-4}
\rho_4 &\;=\;& \big(|\,\phi\,\rangle\langle\,\phi\,|\big)_4 \;=\; \big(\left|\,I_{42}\,\phi_2\,\right\rangle\left\langle\,I_{42}\,\phi_2\,\right|\big)_4 \;=\;
\big(\left|\,I_{40}\,\phi_0\,\right\rangle\left\langle\,I_{40}\,\phi_0\,\right|\big)_4\,.
\end{eqnarray}
Thus the state $\rho_2$ (\ref{rho-phi-2}) is teleported isometrically from $\A_2$ to $\A_4$ satisfying the relation
\begin{equation}\label{isometry relation I42}
I_{42} \;=\; \tilde I_{43} \circ \tilde I_{32} \,.
\end{equation}

Altogether, after two successive measurements, performed first by Alice then by Victor, the incoming state $\rho_0$ (\ref{phi0 at A0}) is teleported isometrically from $\A_0$ to $\A_4\,$, see Fig.~\ref{fig:teleportation with delayed entanglement swapping}, and it satisfies the isometry relation
\begin{equation}\label{isometry relation I04}
I_{40} \;=\; I_{42} \circ I_{20} \;=\; \tilde I_{43} \circ \tilde I_{32} \circ \tilde I_{21} \circ \tilde I_{10} \,,
\end{equation}
which is determined by the entangled states $|\,\psi\,\rangle_{i,i+1}\,$.\\

\begin{figure}
	\centering
	\includegraphics[width=0.50\textwidth]{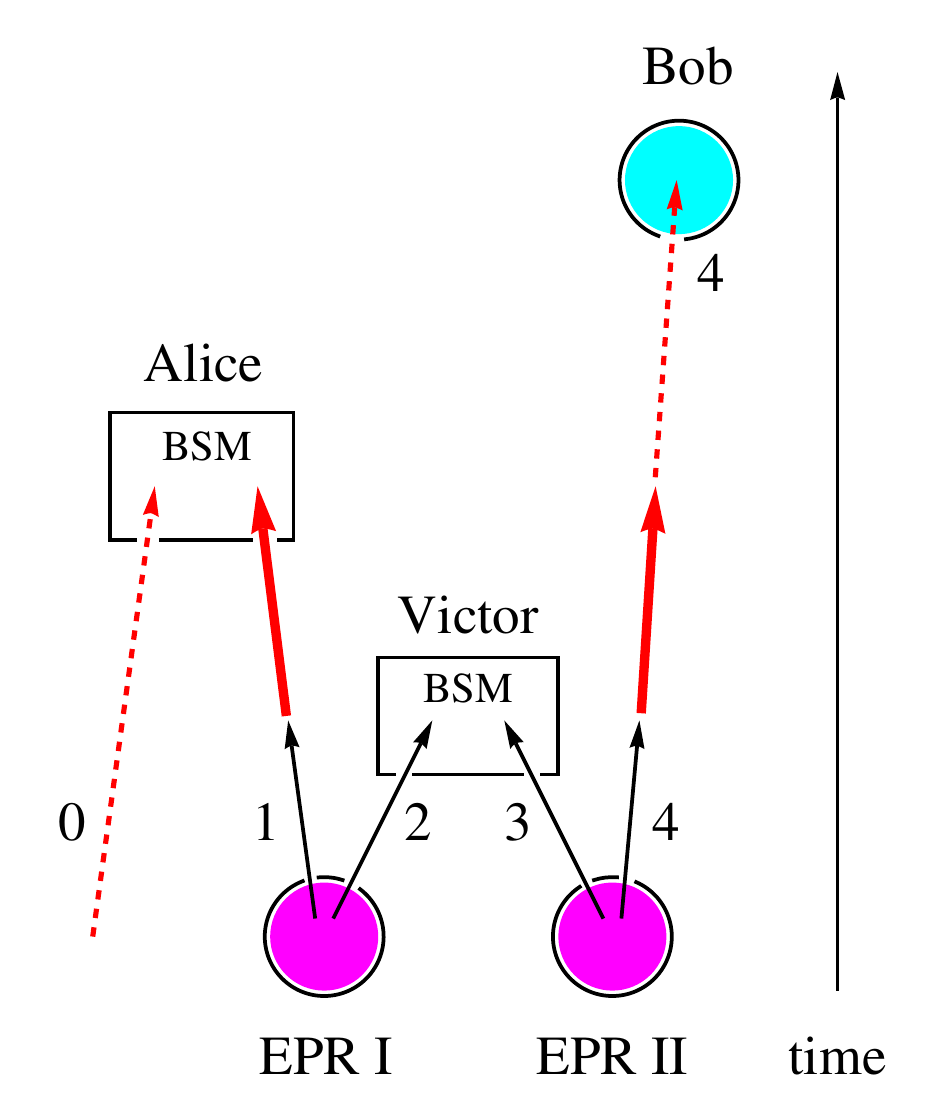}
	\caption{Teleportation with entanglement swapping. There are two EPR sources I and II and we have quantum teleportation of an incoming state at channel 0 to the same outgoing state at channel 4, measured by Bob, with help of entanglement swapping from channels (2,3) to channels (1,4). The order of measurements is reversed as compared to Fig.~\ref{fig:teleportation with delayed entanglement swapping}$\,$, i.e., first Victor performs a Bell state measurement in channels (2,3) and later on Alice in channels (0,1) teleporting the incoming state from 0 to 4$\,$. When comparing this case with the set-up of Fig.~\ref{fig:teleportation with delayed entanglement swapping}$\,$, Bob does not detect any difference in the order of the measurements, again due to the commutativity of the projection operators.}
	\label{fig:teleportation with entanglement swapping}
\end{figure}

Next we consider the reversed order of successive measurements, which corresponds to entanglement swapping, see Fig.~\ref{fig:teleportation with entanglement swapping}$\,$.
The first measurement is $M_{\rm{Victor}} = \{Q_{23}\}\,$, where Victor projects onto some maximally entangled state $\rho_{23} \equiv Q_{23}$ on $\A_2 \otimes \A_3\,$. Then the initial state $\rho_{\rm{initial}}$ (\ref{rho total initially}) changes into
\begin{equation}\label{rho-add0-entangled12-entangled34 first M-Victor}
\rho_{\rm{initial}} \;\;\stackrel{M_{\rm{Victor}}\quad\;\;}{\longrightarrow}\;\;
Q_{23} \,\rho_0 \otimes \rho_{12} \otimes \rho_{34} \, Q_{23} \;=\; \rho_0 \otimes \rho_{14} \otimes \rho_{23} \,,
\end{equation}
with $\rho_{14} \,=\, \big(\left|\,\psi\,\right\rangle\left\langle\,\psi\,\right|\big)_{14}\,$. That means we achieved entanglement swapping from $\A_2 \otimes \A_3$ to $\A_1 \otimes \A_4\,$.

In a second measurement $M_{\rm{Alice}} = \{Q_{01}\}\,$, Alice projects on some entangled state $\rho_{01} \equiv Q_{01}$ on $\A_0 \otimes \A_1\,$. Then the state (\ref{rho-add0-entangled12-entangled34 first M-Victor}) transforms into
\begin{equation}\label{rho-add0-entangled12-entangled34 first M-Victor second M-Alice}
\rho \;=\; \rho_0 \otimes \rho_{14} \otimes \rho_{23} \;\;\stackrel{M_{\rm{Alice}}\quad\;}{\longrightarrow}\;\;
Q_{01} \,\rho_0 \otimes \rho_{14} \otimes \rho_{23}\, Q_{01} \;=\; \rho_{01} \otimes \rho_{23} \otimes \rho_4 \,,
\end{equation}
where $\rho_4$ is given by Eq.~(\ref{rho-phi-4}). Thus we achieve isometric teleportation of the state $\rho_0$ at $\A_0$ to $\rho_4$ at $\A_4\,$, see Fig.~\ref{fig:teleportation with entanglement swapping}$\,$, arriving precisely at the previous result of Eq.~(\ref{rho-add0-entangled12-entangled34 first M-Alice second M-Victor}).

Summarizing, for the reversed case (Fig.~\ref{fig:teleportation with entanglement swapping}) we find that the incoming state $\rho_0$ (\ref{phi0 at A0}) is teleported isometrically from $\A_0$ to $\A_4\,$ and satisfies the same isometry relation (\ref{isometry relation I04}) as before. Therefore Bob's result on $\A_4$ is independent of the order of the successive measurements. The reason is simply that the projection operators -- the measurement operators -- refer to different spaces and therefore commute
\begin{equation}\label{commutator Q01Q23 vanishing}
\left[\,Q_{01}\,,\,Q_{23}\,\right] \;=\; 0 \,.
\end{equation}

\subsection{Noncommutative Projection Operators}\label{noncommutative projection operators}

\begin{figure}
	\centering
	\includegraphics[width=0.50\textwidth]{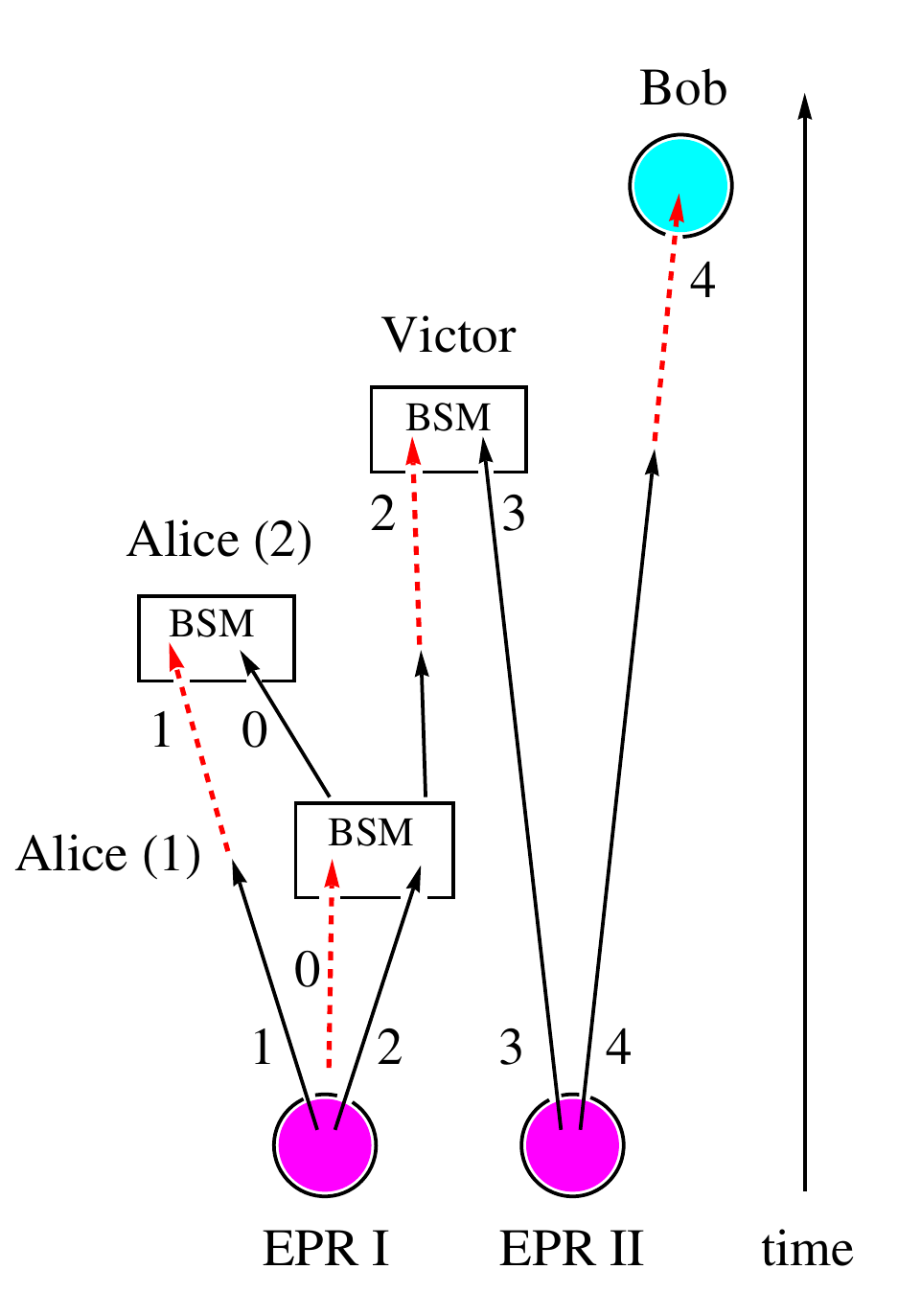}
	\caption{Teleportation with delayed entanglement swapping. There are two EPR sources I and II and we have triple quantum teleportation of an incoming state at channel 0 to the same outgoing state at channel 4, which is measured by Bob. In between there are three Bell state measurements, the first one by Alice (1) in the channels (0,2), teleporting the incoming state from 0 to 1, the second one by Alice (2), teleporting the state from 1 to 2, and the third measurement by Victor in channels (2,3), producing a kind of delayed entanglement swapping, which teleports the state from 2 to 4$\,$.}
	\label{fig:teleportation dependent on delayed entanglement swapping}
\end{figure}

Finally, we construct an experimental set-up, where the results \emph{do} depend on the order of successive measurements, which corresponds to a situation with noncommutative projection operators. In this case, Alice has to measure twice different channels and entanglement swapping is used to teleport an incoming quantum state at Alice's side to Bob, where Bob has now the possibility to examine the measurement sequence corresponding to this noncommutativity of the projection operators, see Fig.~\ref{fig:teleportation dependent on delayed entanglement swapping}$\,$.

Again we start with the initial state $\rho_{\rm{initial}}$ (\ref{rho total initially})$\,$. At the beginning Alice measures twice, first the channel $\A_0 \otimes \A_2$ and afterwards the channel $\A_0 \otimes \A_1\,$. Thus the first BSM is $M^{\,(1)}_{\rm{Alice}} = \{Q_{02}\}$ with the corresponding projection operator $Q_{02} \,=\, \big(\left|\,\psi\,\right\rangle\left\langle\,\psi\,\right|\big)_{02}\,$, Alice projects onto some maximally entangled state $\rho_{02} \equiv Q_{02}$ on $\A_0 \otimes \A_2\,$. Then the initial state changes into
\begin{equation}\label{rho-add0-entangled12-entangled34 first M-Alice02}
\rho_{\rm{initial}} \;\;\stackrel{M^{\,(1)}_{\rm{Alice}}\quad\;}{\longrightarrow}\;\;
Q_{02} \,\rho_0 \otimes \rho_{12} \otimes \rho_{34} \, Q_{02} \;=\; \rho_1 \otimes \rho_{02} \otimes \rho_{34} \,,
\end{equation}
where the state
\begin{eqnarray}\label{rho-phi-1}
\rho_1 &\;=\;& \big(|\,\phi\,\rangle\langle\,\phi\,|\big)_1 \;=\;
\big(\left|\,I_{10}\,\phi_0\,\right\rangle\left\langle\,I_{10}\,\phi_0\,\right|\big)_1
\end{eqnarray}
satisfies the isometry relation
\begin{equation}\label{isometry relation I01}
I_{10} \;=\; \tilde I_{12} \circ \tilde I_{20} \,.
\end{equation}
Thus the incoming state $\rho_0$ (\ref{phi0 at A0}) is teleported isometrically from $\A_0$ to $\A_1\,$, see Fig.~\ref{fig:teleportation dependent on delayed entanglement swapping}$\,$.

In a second measurement $M^{\,(2)}_{\rm{Alice}} = \{Q_{01}\}$ with the corresponding projection operator $Q_{01} \,=\, \big(\left|\,\psi\,\right\rangle\left\langle\,\psi\,\right|\big)_{01}\,$, Alice projects onto a maximally entangled state $\rho_{01} \equiv Q_{01}$ on $\A_0 \otimes \A_1\,$. Then the state (\ref{rho-add0-entangled12-entangled34 first M-Alice02}) transforms into
\begin{equation}\label{rho-add0-entangled12-entangled34 first M-Alice02 second M-Alice01}
\rho \;=\; \rho_1 \otimes \rho_{02} \otimes \rho_{34} \;\;\stackrel{M^{\,(2)}_{\rm{Alice}}\quad\;}{\longrightarrow}\;\;
Q_{01} \,\rho_1 \otimes \rho_{02} \otimes \rho_{34} \, Q_{01} \;=\;\rho_{01} \otimes \rho_2 \otimes \rho_{34} \,.
\end{equation}
Now the state $\rho_2$ is determined by the isometry
\begin{eqnarray}\label{rhoA-phi-2}
\rho_2 &\;=\;& \big(|\,\phi\,\rangle\langle\,\phi\,|\big)_2 \;=\;
\big(\left|\,I_{21}\,\phi_1\,\right\rangle\left\langle\,I_{21}\,\phi_1\,\right|\big)_2 \;=\;
\big(\left|\,I_{20}\,\phi_0\,\right\rangle\left\langle\,I_{20}\,\phi_0\,\right|\big)_2
\end{eqnarray}
and need not be identical to $\rho_2$ (\ref{rho-phi-2}) of the previous case. The reason is that the state $\rho_2$ is the teleportation result of an already teleported state $\rho_1$ and has to satisfy as such a more involved isometry relation. Thus the state $\rho_1$ is again teleported isometrically from $\A_1$ to $\A_2\,$, see Fig.~\ref{fig:teleportation dependent on delayed entanglement swapping}$\,$. Therefore, the state $\rho_2$ obeys the isometry relation
\begin{equation}\label{isometry relation I12}
I_{21} \;=\; \tilde I_{20} \circ \tilde I_{01} \,,
\end{equation}
which is determined by the entangled states $|\,\psi\,\rangle_{01}$ and $|\,\psi\,\rangle_{02}\,$. In relation to the incoming state $\rho_0$ (\ref{phi0 at A0})$\,$, accordingly the state $\rho_2$ (\ref{rhoA-phi-2}) has to satisfy the product relation
\begin{equation}\label{isometry relation I02-A}
I_{20} \;=\; I_{21} \circ I_{10} \;=\;
\tilde I_{20} \circ \tilde I_{01} \circ
\tilde I_{12} \circ \tilde I_{20} \,.
\end{equation}

Next, in a third measurement $M_{\rm{Victor}} = \{Q_{23}\}$ as usual Victor projects onto a maximally entangled state $\rho_{23} \equiv Q_{23}$ on $\A_2 \otimes \A_3\,$. Then the state (\ref{rho-add0-entangled12-entangled34 first M-Alice02 second M-Alice01}) changes into
\begin{equation}\label{rho-add0-entangled12-entangled34 first M-Alice02 second M-Alice01 third Victor23}
\rho \;=\; \rho_{01} \otimes \rho_2 \otimes \rho_{34} \;\;\stackrel{M_{\rm{Victor}}\quad\;\;}{\longrightarrow}\;\;
Q_{23} \,\rho_{01} \otimes \rho_2 \otimes \rho_{34} \, Q_{23} \;=\;\rho_{01} \otimes \rho_{23} \otimes \rho_4 \,,
\end{equation}
where the state $\rho_4$ is defined by
\begin{eqnarray}\label{rhoA-phi-4}
\rho_4 &\;=\;& \big(|\,\phi\,\rangle\langle\,\phi\,|\big)_4 \;=\; \big(\left|\,I_{42}\,\phi_2\,\right\rangle\left\langle\,I_{42}\,\phi_2\,\right|\big)_4 \;=\;
\big(\left|\,I_{40}\,\phi_0\,\right\rangle\left\langle\,I_{40}\,\phi_0\,\right|\big)_4\,.
\end{eqnarray}
That means the state $\rho_2$ (\ref{rhoA-phi-2}) is teleported isometrically from $\A_2$ to $\A_4$ satisfying the relation
\begin{equation}\label{isometry relation I42}
I_{42} \;=\; \tilde I_{43} \circ \tilde I_{32} \,.
\end{equation}

Finally, altogether we obtain teleportation of an incoming state $\rho_0$ (\ref{phi0 at A0}) at $\A_0$ to a state $\rho_4$
(\ref{rhoA-phi-4}) at $\A_4$, see Fig.~\ref{fig:teleportation dependent on delayed entanglement swapping}$\,$, which has to satisfy the isometry relation of three partial teleportations,
\begin{eqnarray}\label{isometry relation I40-1}
I_{40}^{\,1} \;=\; I_{42} \circ I_{21} \circ I_{10}
\;=\; \tilde I_{43} \circ \tilde I_{32} \circ \tilde I_{20} \circ \tilde I_{01} \circ \tilde I_{12} \circ \tilde I_{20}\,,
\end{eqnarray}
where the upper index refers to the first sequence of successive measurements.\\

\begin{figure}
	\centering
	\includegraphics[width=0.50\textwidth]{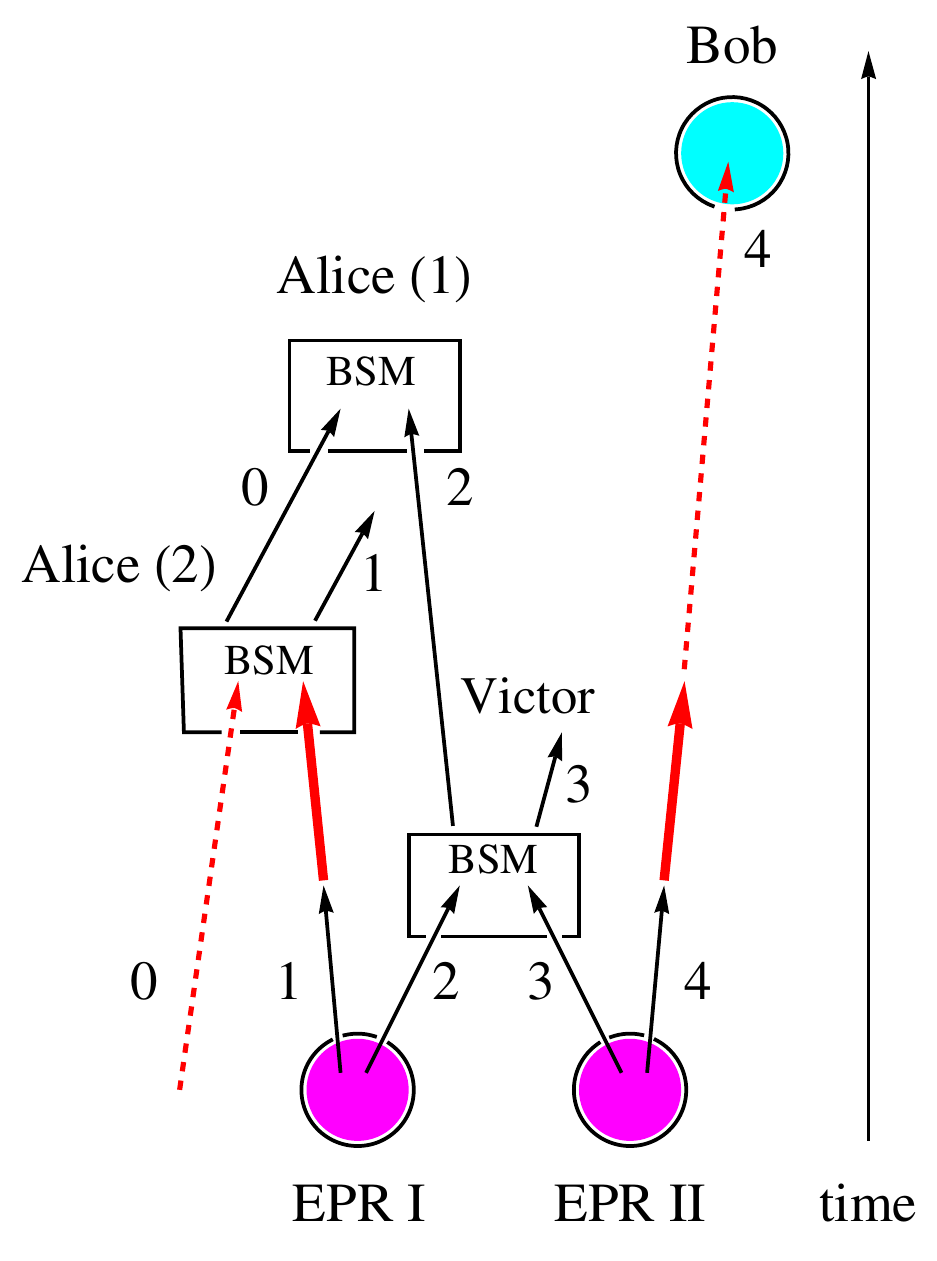}
	\caption{Teleportation with entanglement swapping. There are two EPR sources I and II and we have a reversed order of measurements as compared to Fig.~\ref{fig:teleportation dependent on delayed entanglement swapping}$\,$. First Victor entangles the photons in channels (2,3), i.e., he performs a Bell state measurement, and the remaining two photons in channels (1,4) become instantaneously entangled into the same entangled state. Second Alice (2) performs a Bell state measurement, teleporting the incoming state from 0 to 4, and third Alice (1) entangles the photons in channels (0,2). Channel 4 is finally measured by Bob. When comparing this case with the set-up of Fig.~\ref{fig:teleportation dependent on delayed entanglement swapping}$\,$, Bob \emph{does} now have the possibility to detect the order of the successive measurements due to the noncommutativity of the projection operators.}
	\label{fig:teleportation dependent on entanglement swapping noncomm}
\end{figure}

Considering now the reversed order of successive measurements, which again corresponds to entanglement swapping, we have the following measurement procedure, see Fig.~\ref{fig:teleportation dependent on entanglement swapping noncomm}$\,$. In a first measurement $M_{\rm{Victor}} = \{Q_{23}\}\,$, Victor projects onto some maximally entangled state $\rho_{23} \equiv Q_{23}$ on $\A_2 \otimes \A_3\,$, which produces entanglement swapping from $\A_2 \otimes \A_3$ to $\A_1 \otimes \A_4\,$. In a second one, $M^{\,(2)}_{\rm{Alice}} = \{Q_{01}\}\,$, Alice projects onto a maximally entangled state $\rho_{01} \equiv Q_{01}$ on $\A_0 \otimes \A_1\,$, which causes isometric teleportation of the state $\rho_0$ at $\A_0$ to $\rho_4$ at $\A_4\,$. Finally, in a third measurement $M^{\,(1)}_{\rm{Alice}} = \{Q_{02}\}\,$, Alice projects again onto the entangled state $\rho_{02} \equiv Q_{02}$ on $\A_0 \otimes \A_2\,$, see Fig.~\ref{fig:teleportation dependent on entanglement swapping noncomm}$\,$.

\vspace{0.2cm}

In detail we have
\begin{equation}\label{rho-add0-entangled12-entangled34 reversed first Victor23}
\rho_{\rm{initial}} \;\;\stackrel{M_{\rm{Victor}}\quad\;\;}{\longrightarrow}\;\;
Q_{23} \,\rho_0 \otimes \rho_{12} \otimes \rho_{34} \, Q_{23} \;=\; \rho_0 \otimes \rho_{14} \otimes \rho_{23} \,,
\end{equation}
\begin{equation}\label{rho-add0-entangled12-entangled34 reversed first Victor second M-Alice01}
 \qquad\quad\stackrel{M^{\,(2)}_{\rm{Alice}}\quad\;}{\longrightarrow}\;\;
Q_{01} \,\rho_0 \otimes \rho_{14} \otimes \rho_{23} \, Q_{01} \;=\; \rho_{01} \otimes \rho_{23} \otimes \rho_4 \,,
\end{equation}
\begin{equation}\label{rho-add0-entangled12-entangled34 reversed first Victor second M-Alice01 third Alice01}
 \qquad\quad\stackrel{M^{\,(1)}_{\rm{Alice}}\quad\;}{\longrightarrow}\;\;
Q_{02} \,\rho_{01} \otimes \rho_{23} \otimes \rho_4 \, Q_{02} \;=\; \rho_{02} \otimes \rho_{13} \otimes \rho_4 \,,
\end{equation}
where now the teleported state $\rho_4$ has to satisfy the isometry relation (the upper index refers to the second sequence of successive measurements)
\begin{eqnarray}\label{isometry relation I40-2}
I_{40}^{\,2} \;=\; \tilde I_{41} \circ \tilde I_{10}
\;=\; \tilde I_{43} \circ \tilde I_{32} \circ \tilde I_{21} \circ \tilde I_{10} \;\equiv\; I_{40}\,.
\end{eqnarray}
In this case we have to insert the entanglement swapping into channels (1,4) characterized by $\tilde I_{41}$ (\ref{isometry relation for 1-4}) and the BSM in channels (0,1) given by $\tilde I_{10}\,$. Altogether it coincides with the previous relation (\ref{isometry relation I04}) for $I_{40}\,$.\\

As we can see, in case of noncommutative projection operators, quite generally, the isometry relations (\ref{isometry relation I40-1}) and (\ref{isometry relation I40-2}) for the teleported state $\rho_4$ on Bob's side may differ for the two discussed physical set-ups and therefore need not be identical. For this reason Bob has the possibility to find out the precise order of successive measurements by examining his teleported state. Let us formulate it more precisely.

The difference of the isometries $I_{40}^{\,1}$ and $I_{40}^{\,2}$ for the two set-ups can be traced back to the difference of the following isometries
\begin{eqnarray}\label{isometry I20-1}
I_{20}^{\,1} &\;:=\;& \tilde I_{20} \circ \tilde I_{01} \circ \tilde I_{12} \circ \tilde I_{20}\\
I_{20}^{\,2} &\;:=\;& \tilde I_{21} \circ \tilde I_{10}\,.\label{isometry I20-2}
\end{eqnarray}
Therefore the possibility to detect the noncommutativity of the measurements reduces to the question when do the above isometries coincide, $I_{20}^{\,1} = I_{20}^{\,2}\,$, and when not, $I_{20}^{\,1} \neq I_{20}^{\,2}\,$. Let us investigate two such cases.\\

Firstly, we choose the involved isometries by the following tensor products (entangled states) as defined in Eq.~(\ref{state vectors in teleportation})$\,$:
\begin{eqnarray}\label{isometry tildeI20}
\tilde I_{20} &\;:\;& \quad |\,\varphi_{i} \,\rangle_0 \otimes |\,\varphi_{i} \,\rangle_2 \,,\\\label{isometry tildeI10}
\tilde I_{10} &\;:\;& \quad |\,\varphi_{i} \,\rangle_0 \otimes |\,\varphi_{i} \,\rangle_1 \,,\\
\tilde I_{21} &\;:\;& \quad |\,\varphi_{i} \,\rangle_1 \otimes |\,\varphi_{i} \,\rangle_2 \,.\label{isometry tildeI21}
\end{eqnarray}
Then the isometries $I_{20}^{\,1}\,, I_{20}^{\,2}$ of the two set-ups correspond to the two chains of mappings
\begin{eqnarray}\label{isometry I20-1 mapping chain}
I_{20}^{\,1} &\;:\;& \quad |\,\varphi_{i} \,\rangle_2 \;\;\stackrel{\;\tilde I_{20}}{\longleftarrow}\; |\,\varphi_{i} \,\rangle_0 \;\;\stackrel{\;\tilde I_{01}}{\longleftarrow}\; |\,\varphi_{i} \,\rangle_1 \;\;\stackrel{\;\tilde I_{12}}{\longleftarrow}\; |\,\varphi_{i} \,\rangle_2 \;\;\stackrel{\;\tilde I_{20}}{\longleftarrow}\; |\,\varphi_{i} \,\rangle_0\\
I_{20}^{\,2} &\;:\;& \quad |\,\varphi_{i} \,\rangle_2 \;\;\stackrel{\;\tilde I_{21}}{\longleftarrow}\; |\,\varphi_{i} \,\rangle_1 \;\;\stackrel{\;\tilde I_{10}}{\longleftarrow}\; |\,\varphi_{i} \,\rangle_0\;.\label{isometry I20-2 mapping chain}
\end{eqnarray}
In both cases we have the identical mapping $|\,\varphi_{i} \,\rangle_2 \;\leftarrow\; |\,\varphi_{i} \,\rangle_0$ and consequently the isometries agree,
\begin{equation}\label{I20-1 equal I20-2}
I_{20}^{\,1} \;=\; I_{20}^{\,2}\,,
\end{equation}
and we are not able to detect the order of successive measurements.\\

Secondly, if we choose on the other hand the isometry $\tilde I_{20}$ like
\begin{eqnarray}\label{isometry tildeI20prime}
\tilde I_{20}^{\,'} &\;:\;& \quad |\,\varphi_{i} \,\rangle_0 \otimes |\,\varphi_{i + 1} \,\rangle_2 \;,
\end{eqnarray}
and the other isometries (\ref{isometry tildeI10}) and (\ref{isometry tildeI21}) remain the same, then we obtain a different chain of mappings for the first set-up
\begin{eqnarray}\label{isometry I20-1prime mapping chain}
I_{20}^{\,'\,1} &\;:\;& \quad |\,\varphi_{i + 2} \,\rangle_2 \;\;\stackrel{\;\tilde I_{20}^{\,'}}{\longleftarrow}\; |\,\varphi_{i + 1} \,\rangle_0 \;\;\stackrel{\;\tilde I_{01}}{\longleftarrow}\; |\,\varphi_{i + 1} \,\rangle_1 \;\;\stackrel{\;\tilde I_{12}}{\longleftarrow}\; |\,\varphi_{i + 1} \,\rangle_2 \;\;\stackrel{\;\tilde I_{20}^{\,'}}{\longleftarrow}\; |\,\varphi_{i} \,\rangle_0\;.
\end{eqnarray}
The isometry $I_{20}^{\,'\,1}$ defines the mapping $|\,\varphi_{i + 2} \,\rangle_2 \;\leftarrow\; |\,\varphi_{i} \,\rangle_0\,$, where the new states are determined by unitary transformations $U$ of the originally chosen ones $|\,\varphi_{i + 1} \,\rangle_0 \,=\, U_0\, |\,\varphi_{i} \,\rangle_0$ and $|\,\varphi_{i + 2} \,\rangle_0 \,=\, U_0^{\,2}\, |\,\varphi_{i} \,\rangle_0\,$. In this case (for this simple example we choose $d > 2\,$) we will have in contrast to above relation (\ref{I20-1 equal I20-2})
\begin{equation}\label{I20-1prime notequal I20-2}
I_{20}^{\,'\,1} \;\neq\; I_{20}^{\,2}\,,
\end{equation}
and we are now able to detect the noncommutativity of the operations with the two different physical set-ups.\\

More generally, we can characterize an entangled state by a fixed isometry together with an appropriate unitary transformation. Then Alice has to choose in her first measurement $M^{\,(1)}_{\rm{Alice}} = \{Q_{20}^{\,'}\}$ (see Fig.~\ref{fig:teleportation dependent on delayed entanglement swapping}) the projection operator $Q_{20}^{\,'} \,=\, \big(\left|\,\psi^{\,'}\,\right\rangle\left\langle\,\psi^{\,'}\,\right|\big)_{20}$ with $|\,\psi^{\,'}\,\rangle_{20} \,=\, \mathds{1}_{2} \otimes U_0 \,|\,\psi\,\rangle_{20}\,$, or in terms of isometries
\begin{equation}\label{I20prime equal I20-2 times U}
\tilde I_{20}^{\,'} \;=\; \tilde I_{20} \,U_0\,.
\end{equation}
Inserting the unitarily transformed isometry (\ref{I20prime equal I20-2 times U}) into the composition (\ref{isometry I20-1})$\,$, we get
\begin{eqnarray}\label{isometry I20-1prime U-transformed}
I_{20}^{\,'\,1} &\;:=\;& \tilde I_{20}\,U_0 \circ \tilde I_{01} \circ \tilde I_{12} \circ \tilde I_{20}\,U_0\,.
\end{eqnarray}
Commuting $U$ through all the other isometries (recall Eq.~(\ref{isometry32}))$\,$, we find
\begin{equation}\label{isometry I20-1prime UstarU transformed}
I_{20}^{\,'\,1} \;=\; U^{\ast}_2 \,U_2 \,I_{20}^{\,1}\,,
\end{equation}
which gives $I_{20}^{\,'\,1} \,\neq\, I_{20}^{\,1} \,=\, I_{20}^{\,2}$ as we obtained before in Eq.~(\ref{I20-1prime notequal I20-2}) since in general $U^{\ast} \,U \neq \mathds{1}$ (for the complex conjugate of an operator the unitary relation does not necessarily hold).

Thus the second choice (\ref{isometry tildeI20prime}) of the isometry means that Alice just has to choose her projection operator $Q_{20}^{\,'}$ appropriately (see Fig.~\ref{fig:teleportation dependent on delayed entanglement swapping})$\,$, then Bob will be able to verify the noncommutativity of the measurements with respect to time ordering.

We also want to draw attention to the fact that it is quite important to have a $\mathds{C}$-valued vector space in quantum mechanics and not a $\mathds{R}$-valued one in order to find $U^{\ast} \neq U^{\dagger}\,$, what we need in the above analysis.\\

Considering finally the involved projection operators, quite generally, they do not commute$\,$:
\begin{equation}\label{noncommuting projection operators}
Q_{02} \circ Q_{01} \circ Q_{23} \;\neq\; Q_{23} \circ Q_{01} \circ Q_{02} \,,
\end{equation}
since obviously Eqs.(\ref{rho-add0-entangled12-entangled34 first M-Alice02 second M-Alice01 third Victor23}) and (\ref{rho-add0-entangled12-entangled34 reversed first Victor second M-Alice01 third Alice01}) define different states. The appropriate choice is necessary to observe the noncommutativity for the reduction of the state to Bob's space $\Ha_4\,$.

What we further notice is that in the second set-up for the reversed order of successive measurements (see Fig.~\ref{fig:teleportation dependent on entanglement swapping noncomm}) the third measurement $M^{\,(1)}_{\rm{Alice}} = \{Q_{02}\}$ is actually not needed and it is sufficient to consider just the case of Fig.~\ref{fig:teleportation with entanglement swapping}$\,$.

\section{Conclusion} \label{sec:conclusion}


We consider a sequence of measurements useful for quantum teleportation and entanglement swapping in different orders. The analysis is formulated in terms of isometries that turn out to be a powerful tool for allowing us to work in all dimensions and to concentrate on the essential features of the phenomena.

Based on the interpretation that every measurement corresponds to a collapse of the quantum state, the quantum history built by a sequence of projection operators is not independent of the ordering of the measurements due to the noncommutativity of the projection operators. If, however, the measurements correspond to algebraically independent systems, these projection operators commute and the time ordering is irrelevant as it was observed for delayed-choice entanglement swapping.

If, on the other hand, it is possible to carry out measurements on overlapping algebras as it is the case for a possibly repeated action on a photon, or spin-$\frac{1}{2}$ particle, then we are able to arrange the measurements in such a way that we can observe the noncommutativity with respect to time ordering with certainty.

For the direct comparison of our analysis with photon experiments, a definite polarization quantum state must be chosen for both the EPR sources and the Bell state measurements. But which one will depend on the experimental possibility available and can only be decided by the experimenter.

Last but not least, in our whole analysis we tacitly assumed that after a definite Bell state measurement the photons still remain in this definite Bell state and are available for further measurements. However, present technologies do not offer these possibilities. We consider it as a challenge for experimentalists, and we do hope that with advanced technologies it will become possible to detect the noncommutativity of successive measurements in Nature.

\begin{acknowledgments}

We would like to thank Philip Walther and Anton Zeilinger for helpful discussions and Gerhard Ecker for the careful reading of the manuscript.

\end{acknowledgments}


\begin{thebibliography}{100}

\bibitem{Ma-Zeilinger}
X.-S. Ma, S. Zotter, J. Kofler, R. Ursin, T. Jennewein, C. Brukner, and A. Zeilinger, Nature Physics \textbf{8}, 480 (2012)

\bibitem{Peres00}
A. Peres, Journ. Mod. Optics \textbf{47}, 139 (2000)

\bibitem{vonNeumann}
J. von Neumann, \emph{Mathematische Grundlagen der Quantenmechanik}, Springer 1996

\bibitem{Gell-Mann_Hartle}
M. Gell-Mann and J. B. Hartle, Phys. Rev. D \textbf{47}, 3345 (1993)

\bibitem{Griffiths}
R. B. Griffiths, Phys. Rev. A \textbf{54}, 2759 (1996), Phys. Rev. A \textbf{57}, 1604 (1998)

\bibitem{Uhlmann01}
A. Uhlmann, "Antilinearity in bipartite quantum systems and imperfect quantum teleportation", in \emph{Quantum probability and white noise analysis}, Vol. XV, Proceedings of the conference \emph{Quantum probability and infinite dimensional analysis}, W. Freudenberg (ed.), World Scientific 2001, p. 255

\bibitem{TBKN}
W. Thirring, R.A. Bertlmann, P. K\" ohler, and H. Narnhofer, Eur. Phys. J. D \textbf{64}, 181 (2011)

\bibitem{bell1964}
J. S. Bell, Physics \textbf{1}, 195 (1964)

\bibitem{bell-book}
J.S. Bell, \textit{Speakable and unspeakable of quantum mechanics}, Cambridge University Press (1987)

\bibitem{horodecki96}
M. Horodecki, P. Horodecki, and R. Horodecki, Phys. Lett. A \textbf{223}, 1 (1996)

\bibitem{terhal00}
B. M. Terhal, Phys. Lett. A \textbf{271}, 319 (2000)

\bibitem{bertlmann-narnhofer-thirring02}
R. A. Bertlmann, H. Narnhofer, and W. Thirring, Phys. Rev. A \textbf{66}, 032319 (2002)

\bibitem{bruss02}
D. Bru\ss, J. Math. Phys. \textbf{43}, 4237 (2002)

\bibitem{BBCJPW}
C. H. Bennett, G. Brassard, C. Cr\'epeau, R. Jozsa, A. Peres, and W.K. Wootters, Phys. Rev. Lett. \textbf{70}, 1895 (1993)

\bibitem{bouwmeester-pan-zeilinger-etal}
D. Bouwmeester, J. W. Pan, K. Mattle, M. Eibl, H. Weinfurter, and A. Zeilinger, Nature \textbf{390}, 575 (1997)

\bibitem{ursin-zeilinger-etal}
R. Ursin, T. Jennewein, M. Aspelmeyer, R. Kaltenbaek, M. Lindenthal, P. Walther, and A. Zeilinger, Nature \textbf{430}, 849 (2004)

\bibitem{EPR}
A. Einstein, B. Podolsky, and N. Rosen, Phys. Rev. A \textbf{47}, 777 (1935)

\bibitem{zukowski-zeilinger-horne-ekert}
M. \.Zukowski, A. Zeilinger, M. Horne, and A. Ekert, Phys. Rev. Lett. \textbf{71}, 4287 (1993)

\bibitem{pan-bouwmeester-weinfurter-zeilinger}
J.-W. Pan, D. Bouwmeester, H. Weinfurter, and A. Zeilinger, Phys. Rev. Lett. \textbf{80}, 3891 (1998)

\bibitem{bertlmann-zeilinger02}
R. A. Bertlmann and A. Zeilinger (eds.), \emph{Quantum [Un]speakables}, Springer 2002


\end{thebibliography}
\end{document}